\documentclass[12pt]{article}
\usepackage[a4paper]{geometry}	

\usepackage{fancyhdr}
\usepackage[version=4]{mhchem}
\usepackage{graphicx} 			
\usepackage{indentfirst}
\usepackage{longtable}
\usepackage{siunitx}
\usepackage{subfigure}
\usepackage{chemfig}
\usepackage{fancyhdr}			
\usepackage{lmodern}			
\usepackage{amsmath}			
\usepackage[hidelinks]{hyperref}

\usepackage{threeparttable}	
\usepackage{url}  
\usepackage{booktabs}			
\usepackage{listings}
\usepackage{xcolor}
\usepackage{cite}

\usepackage{listings}
\usepackage{xcolor}

\graphicspath{{figures/}}		

\begin{document}
	\title{Maze-Bubble Pattern Magnetic Domain Simulation \\Based on the Lengyel-Epstein Model}
	\author{Yufei Bai \\Department of Mathematics\\ University of Science and Technology of China\\ 96 Jinzhai Road, Hefei 230026, Anhui, P. R. China}
	\date{}
	\maketitle
	\begin{abstract}
		This study is based on the Lengyel-Epstein (LE) model, governed by a system of nonlinear partial differential equations, to simulate the maze-bubble pattern magnetic domains in magnetic thin films with perpendicular magnetic anisotropy (PMA). Through numerical simulations, we successfully reproduce the maze, bubble, and intermediate-state magnetic domains observed in PMA multilayer films under the influence of material thickness and external magnetic fields. The topological structures of the magnetic domains shown in the simulation closely resemble those observed under a microscope, demonstrating the effectiveness of the LE model in simulating changes in the magnetic domain topology of magnetic thin films. This study also innovatively applies the concept of reaction-diffusion, commonly used in biochemistry, by drawing an analogy to electromagnetism. This approach holds significant implications for the study of magnetic domains.
	\end{abstract}
	
	
	\section{Introduction}
	
	\subsection{Background}
	In ferromagnetic materials, there is a strong exchange coupling interaction between adjacent atoms. In the absence of an external magnetic field, their spin magnetic moments (the magnetic moments generated by atomic spin motion) tend to align spontaneously within small regions, forming regions of spontaneous magnetization, known as magnetic domains\cite{1, 2}. Magnetic domains are the fundamental units in magnetic materials, aligning spontaneously to minimize total magnetic energy and forming structures like maze and bubble patterns. Under certain conditions, magnetic domains can undergo a transition from one form to another. Studying these phase transitions not only reveals the microscopic evolution of magnetic materials but also provides theoretical support for improving the performance of magnetic storage devices.
	\newpage
	
	The Lengyel-Epstein model (LE model) is a classical chemical reaction-diffusion model initially used to describe the formation of spatiotemporal patterns in biochemical systems. The main feature of this model is that it not only generates spatially periodic structures, such as stripes or spot patterns, but also explains the pattern formation mechanisms widely observed in biological systems, with applications in chemistry, ecology, and even materials science. Due to its broad applicability and ability to reflect spontaneous structural generation processes, the LE model has become a fundamental theoretical tool for studying various natural phenomena.
	
	\subsection{Research purpose}
	
	Since the process of magnetic domain formation is complex and highly random, there has been limited simulation research on magnetic domain shapes. In this study, we innovatively apply the concept of reaction-diffusion from chemistry and biology to electromagnetism, using a system of nonlinear partial differential equations based on the LE model to simulate the maze-bubble pattern of magnetic domains in thin films with perpendicular magnetic anisotropy (PMA). By successfully simulating the maze-bubble pattern, we contribute to a deeper understanding of the formation and evolution of magnetic domain structures, and promote the development of novel magnetic storage technologies.
	
	\subsection{Organization}
	
	In the following discussion, Section 2, Preliminary, provides a detailed introduction to the preparatory knowledge on magnetic domains and the LE model used in this study. Section 3, based on the knowledge outlined in Section 2, offers a new perspective on understanding the formation of magnetic domains. Section 4, based on the perspective introduced in Section 3, presents a simulation of remanent magnetic domains in PMA films. Section 5 extends this research by introducing the cutting-edge magnetic storage device, the magnetic skyrmion, and proposing a new hypothesis based on the analogy between Turing patterns and magnetic domain patterns. Finally, Section 6 concludes the study with a summary and outlook.
	
	\section{Preliminary}
	
	In this section, we will primarily introduce the knowledge related to the simulation object of this experiment and the LE model.
	\subsection{Magnetic domain patterns in ferromagnetic films}
	
	In the absence of an external field, the arrangement of magnetic domains is often highly random, as illustrated in Figure 1(a). When a ferromagnetic material is exposed to an external magnetic field, the internal magnetic domains gradually realign along the direction of the applied field. As the external magnetic field increases to a certain saturation point, nearly all magnetic domains align with the direction of the applied field, and the material reaches a saturated magnetization state, as illustrated in Figure 1(b). However, when the external magnetic field is removed, not all magnetic domains will fully randomize or return to their initial demagnetized state. Instead, some domains will retain the magnetization direction previously influenced by the external field, leading the material to maintain a certain magnetization even in the absence of an external field. This is known as the remanent magnetization state\cite{3}.

	\begin{figure}[htbp]
		\centering
		\subfigure[The magnetic domain arrangement occurs in a disordered state without an external field.]{
			\includegraphics[width=0.41\textwidth]{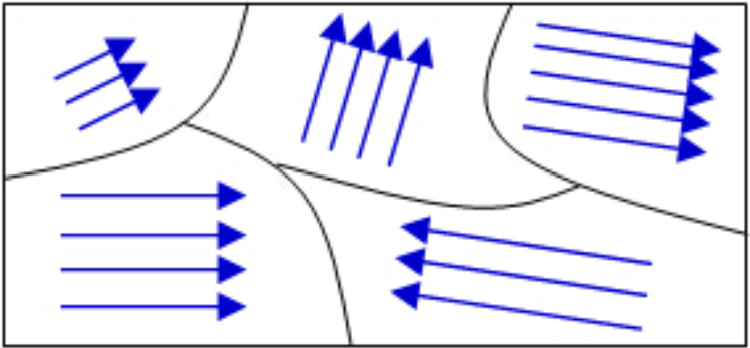}  
		}
		\hspace{0.05\textwidth}  
		\subfigure[After applying an external field, the magnetic moments inside the domain tend to align with the direction of the field.]{
			\includegraphics[width=0.4\textwidth]{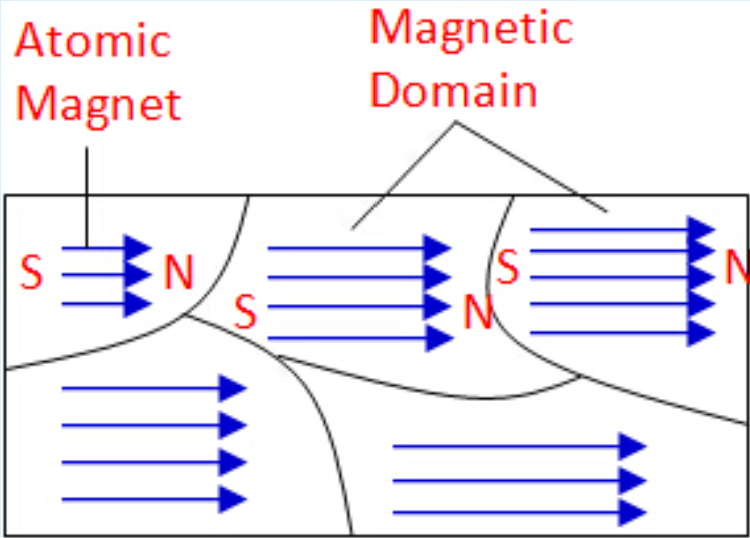}  
		}
		\caption{Schematic of a collection of magnetic domains in a ferromagnetic material\cite{F1}.}
		\label{fig:side_by_side}
	\end{figure}
	
	The topological structure in the remanent magnetization state mainly depends on the thickness of the thin film and the magnetic history, specifically the magnitude of the applied magnetic field during the magnetization process\cite{4}.
	In recent years, magnetic storage has developed rapidly. Studying the topological structure of magnetic materials in the remanent magnetization state helps optimize thin film material performance, improving storage density and stability. 
	
	\subsection{Magnetic domain patterns in thin films with perpendicular magnetic anisotropy}
	
	In thin films with perpendicular magnetic anisotropy (PMA), the magnetization direction of the magnetic material tends to align along the direction perpendicular to the surface of the material. Since the 2000s, PMA thin films have attracted increasing attention because, compared to in-plane magnetization, perpendicular magnetization can significantly reduce the size of magnetic domains, which is crucial for improving magnetic recording density\cite{5}. This is especially advantageous in applications involving high-density data storage. 
	\begin{figure}[htbp]
		\centering
		\subfigure[Maze pattern]{
			\includegraphics[width=0.145\textwidth]{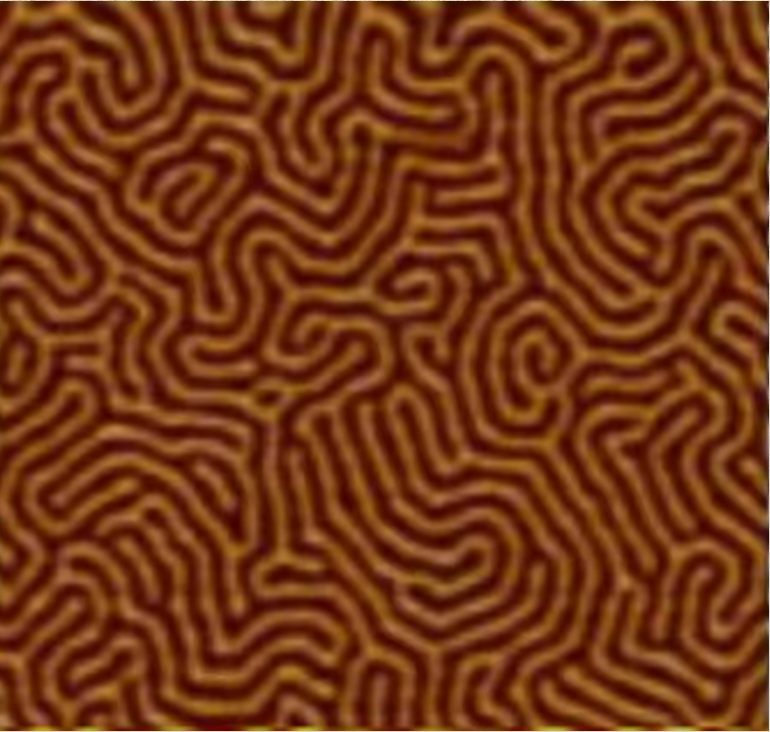}  
		}
		\hspace{0.05\textwidth}  
		\subfigure[Bubble pattern]{
			\includegraphics[width=0.14\textwidth]{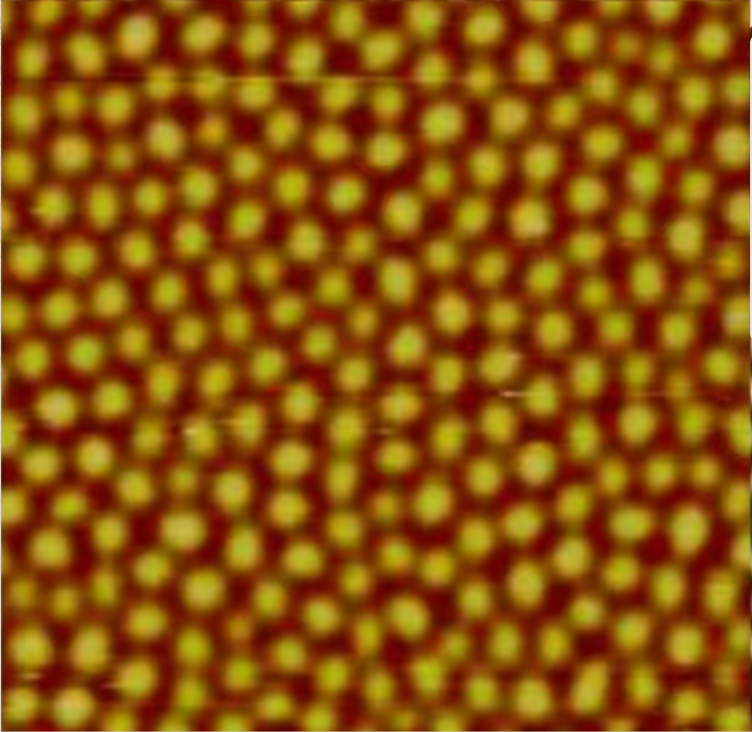}  
		}
		\hspace{0.05\textwidth} 
		\subfigure[Maze-Bubble pattern]{
			\includegraphics[width=0.14\textwidth]{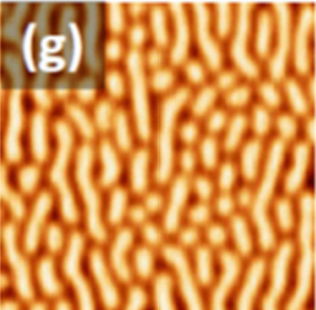}  
		}
		\caption{Remanent magnetic domain observed in
			$[Co/Pt]_N$ multilayers. Extracted from \cite{6}.
		}
		\label{fig:side_by_side1}
	\end{figure}
	
	In smooth PMA thin films, magnetic domains can form various shapes and topological structures, such as stripes, bubbles, and maze-like patterns \cite{7,8,9}. The maze-bubble pattern is a typical magnetic domain structure transition mode in magnetic thin films. In the maze pattern, the magnetic domains exhibit irregularly twisted stripe-like structures, while the bubble pattern consists of independently and uniformly distributed small circular magnetic domains, which exhibit higher stability. This pattern is widely used in high-density magnetic storage devices \cite{6}. Detailed images of the magnetic domain structures can be obtained using Magnetic Force Microscopy (MFM), as illustrated in Figure 2.

	\subsection{Turing pattern}
	
	In 1952, the famous British mathematician Alan Turing turned his attention to the field of biology and successfully explained the principles behind the formation of certain surface patterns in organisms using a reaction-diffusion system \cite{21}.
	\begin{align}
		\frac{\partial u}{\partial t} &= c \Delta u + f(u,v), \\
		\frac{\partial v}{\partial t} &= d \Delta u + g(u,v).
	\end{align}
	
	In the system of equations above, 
	$u$ and $v$ represent two different reactants, and the concentration of these two reactants changes over time due to their diffusion (represented by the Laplace operator) and the coupled reactions (functions 
	$f$ and $g$). The patterns produced by this system are called Turing patterns.
	
	Turing's reaction-diffusion model has attracted significant attention since its introduction and has been extensively developed in theory. However, observing Turing patterns in real chemical or biological systems has been extremely difficult. It was not until the early 1990s that De Kepper and others, through the study of the chlorite-iodide-malic acid (CIMA) reaction, obtained the first experimental evidence of Turing patterns\cite{11}. Since the CIMA reaction involves five reactants, the mathematical analysis of the system became more complex. So, Lengyel and Epstein simplified the model and proposed a two-dimensional version, known as the Lengyel-Epstein model (LE model)\cite{12,13,14}.
	
	\subsection{LE model}
	
	The system of equations in the LE model consists of a nonlinear biased diffusion system formed by two equations:
	\begin{align}
		\frac{\partial U}{\partial t} = \Delta U + a - U - \frac{4UV}{1 + U^2},
		\end {align} 
		\begin{align}
			\frac{\partial V}{\partial t} = \sigma \left[ c \Delta V + b \left( U - \frac{UV}{1 + U^2} \right) \right].
		\end{align}
		
		Here, \( U \) and \( V \) represent the concentrations of two reactants, \( \sigma, a, b, c\) are parameters that control the reaction rates and diffusion properties, and \( \Delta \) represents the Laplace operator, which is used to describe the diffusion effects of each substance.
		
		The operation of the entire system is based on the interaction between reaction and diffusion. The reaction term changes the concentrations of the variables through their local interactions, while the diffusion term drives the distribution of substances across the global space. Under appropriate parameter conditions, this reaction-diffusion process can lead to the formation of periodic or complex patterns in the system\cite{15,a,b}.
		
		\section{A New Perspective on Magnetic Domains}
		
		In this section, we will understand the formation of magnetic domains from a completely new perspective. By drawing an analogy between the concepts of magnetic domains and the LE model, we will innovatively apply the reaction-diffusion model to the field of electromagnetism.
		
		The table below lists the similarities between the LE model and magnetic domains, which serve as the starting point for our analogy between the LE model and magnetic domains.
		\begin{table}[ht]
			\centering
			\begin{tabular}{|p{8.5cm}|p{8.5cm}|}
				\hline
				Magnetic Domain & Lengyel-Epstein Model\\
				\hline
				Complex self-organization phenomenon& Complex self-organization process  \\
				\hline
				Formation of remanent magnetization is \newline equilibrium - non-equilibrium - equilibrium & Reaction-diffusion process is \newline equilibrium - non-equilibrium - equilibrium \\
				\hline
				
				Asymmetric image & Asymmetric image \\
				\hline
			\end{tabular}
		\end{table}
		
		By analogy, we can observe many similarities between magnetic domains and the LE model. Therefore, we can hypothesize—viewing magnetic domains from the perspective of reaction and diffusion, and considering their formation as a process where two reactants interact and diffuse. With this hypothesis, the simulation results can be qualitatively compared with the concepts in the LE model.
		
		\begin{table}[ht]
			\centering
			\begin{tabular}{|p{8.5cm}|p{8.5cm}|}
				\hline
				Remanent Magnetic Domain & Lengyel-Epstein Model\\
				\hline
				The ability of magnetic moments to align \newline perpendicular to the inward/outward direction& The concentration of reactant 1, U \newline and the concentration of reactant 2, V \\
				\hline
				When the external field is saturated, \newline the direction of the magnetic moments is entirely perpendicular to the inward direction. & At the initial moment, U = 1 and V = 0. \\
				\hline
				
				The magnetic domain structure becomes complex and random after the external field is removed& Apply random perturbations to the initial steady state \\
				\hline
				Coupling interaction between adjacent atoms in the perpendicular direction & 
				The reaction terms of U and V \\
				\hline
				Coupling interaction between adjacent atoms in the horizontal direction & The diffusion terms of U and V
				\\
				\hline
				Applied external field, film thickness, and other influencing factors & Parameters in the equation \\
				\hline
			\end{tabular}
		\end{table} 
		\section{The Simulation Process}
		
		Building on the perspective introduced in Section 3, this section presents a simulation of remanent magnetic domains in PMA films, including both the simulation of static magnetic domain maze-bubble patterns and the transformation of domain shapes under varying material thickness and external field strength.
		\subsection{Establish the simulation model}
		
		From this new perspective on magnetic domains, we can begin the simulation of magnetic domains.
		\subsubsection*{Define variables}
		
		Consider a PMA (Perpendicular Magnetic Anisotropy) square thin film with a nanoscale thickness. A Cartesian coordinate system is established on its horizontal plane, where the magnetic moments of all atoms in the vertical direction at each point in the plane are treated as a whole. Since it is a PMA thin film, the direction of the overall magnetic moment is either perpendicular outward or perpendicular inward. Therefore, the magnetic moment vector can be treated as a scalar $M$, where 
		$M=M_1$-$M_2$, with $M_1$ representing the magnitude of the magnetic moment pointing outward, and $M_2$
		representing the magnitude of the magnetic moment pointing inward.
		If $M_1 >M_2$, the magnetic moment direction at that point is perpendicular outward; if 
		$M_1 <M_2$, the magnetic moment direction at that point is perpendicular inward. 
		
		Consider a unit magnetic moment vector $m$, whose time evolution is governed by the Landau-Lifshitz-Gilbert (LLG) equation.
		\begin{align}
			\dot{m}(r, t) = -\gamma m(r, t) \times H_{\text{eff}} + \alpha m(r, t) \times \dot{m}(r, t).
		\end{align}
		Here, $m$ represents a macrospin, and when the effective field is defined as $H_{\text{eff}} = A \nabla^2 m$. The spin wave phenomenon induced by short-range exchange interactions can be observed in a local region, as shown in Figure 3. (Note: 
		$A$ is the exchange stiffness coefficient.)
		\begin{figure}[htbp]
			\centering
			\begin{minipage}{\textwidth}
				\centering
				\includegraphics[width=0.8\textwidth]{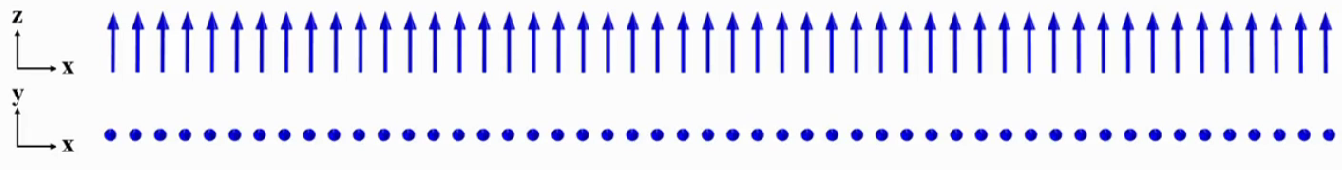} 
				
			\end{minipage}
			
			\vspace{1em} 
			
			\begin{minipage}{\textwidth}
				\centering
				\includegraphics[width=0.8\textwidth]{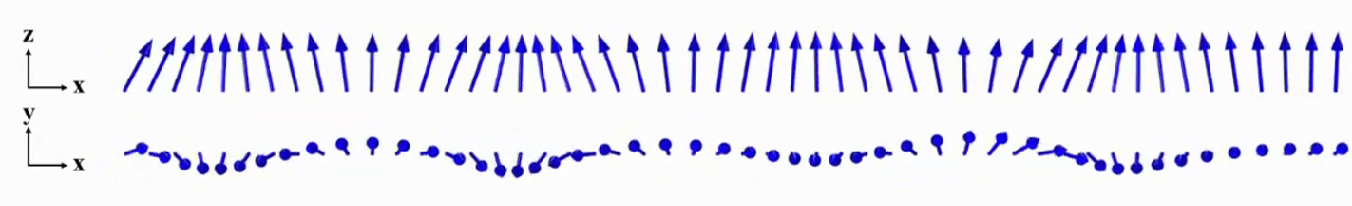} 
				\caption{Short-range interaction-induced spin wave phenomenon in micromagnetic simulation \cite{spin}}
			\end{minipage}
		\end{figure}
		
		Therefore, we approximate the exchange coupling between each point in the plane and its neighboring points as being divided into exchange couplings in the horizontal and vertical directions. For the vertical exchange coupling, a function related to \( M_1 \) and \( M_2 \) is used. For the horizontal exchange coupling, inspired by the spin wave phenomenon, the Laplacian operator of \( M_1 \) and \( M_2 \) is used to represent it.
		
		\subsubsection*{Define the initial state}
		
		The external magnetic field is set to be perpendicular to the material surface, directed outward. As the magnetic field intensity increases, the direction of the magnetic moments within the square region gradually aligns to be outward perpendicular to the material surface, which is defined as the initial state. At $t$ = 0, the magnetic moment at each point in the region is $M_1$ = 1 and $M_2$ = 0. After that, a random disturbance is applied to the region to simulate the condition where the external field has not yet reached saturation strength, causing the magnetic moments in the region to not all align outward. This disturbance breaks the balance of the initial state, causing $M_1$ and $ M_2$ to begin to change.
		\subsubsection*{Establish the core system of equations}
		
		Subsequently, we improve upon the LE model and establish the core system of equations.
		\begin{align}
			\frac{\partial M_1}{\partial t} &= \Delta M_1 + a - M_1 - \frac{w M_1 M_2}{1 + M_1^2}, \\
			\frac{\partial M_2}{\partial t} &= \sigma \left[ c \Delta M_2 + b \left( M_1 - \frac{M_1 M_2}{1 + M_1^2} \right) \right].
		\end{align}
		Here, $M_1$ represents the magnitude of the magnetic moment pointing outward, $M_2$ represents the magnitude of the magnetic moment pointing inward, and $\sigma, a, b, c, w$ are parameters that control the whole system. (Note: This model is a modification of the LE model, where the fixed constant 4 in the original model has been changed to a variable parameter. This change has a significant impact on the simulation in Section 4.3.)
		\subsection{Preliminary simulation results}

		\begin{figure}[ht]
			\centering
			\includegraphics[width=0.66\linewidth]{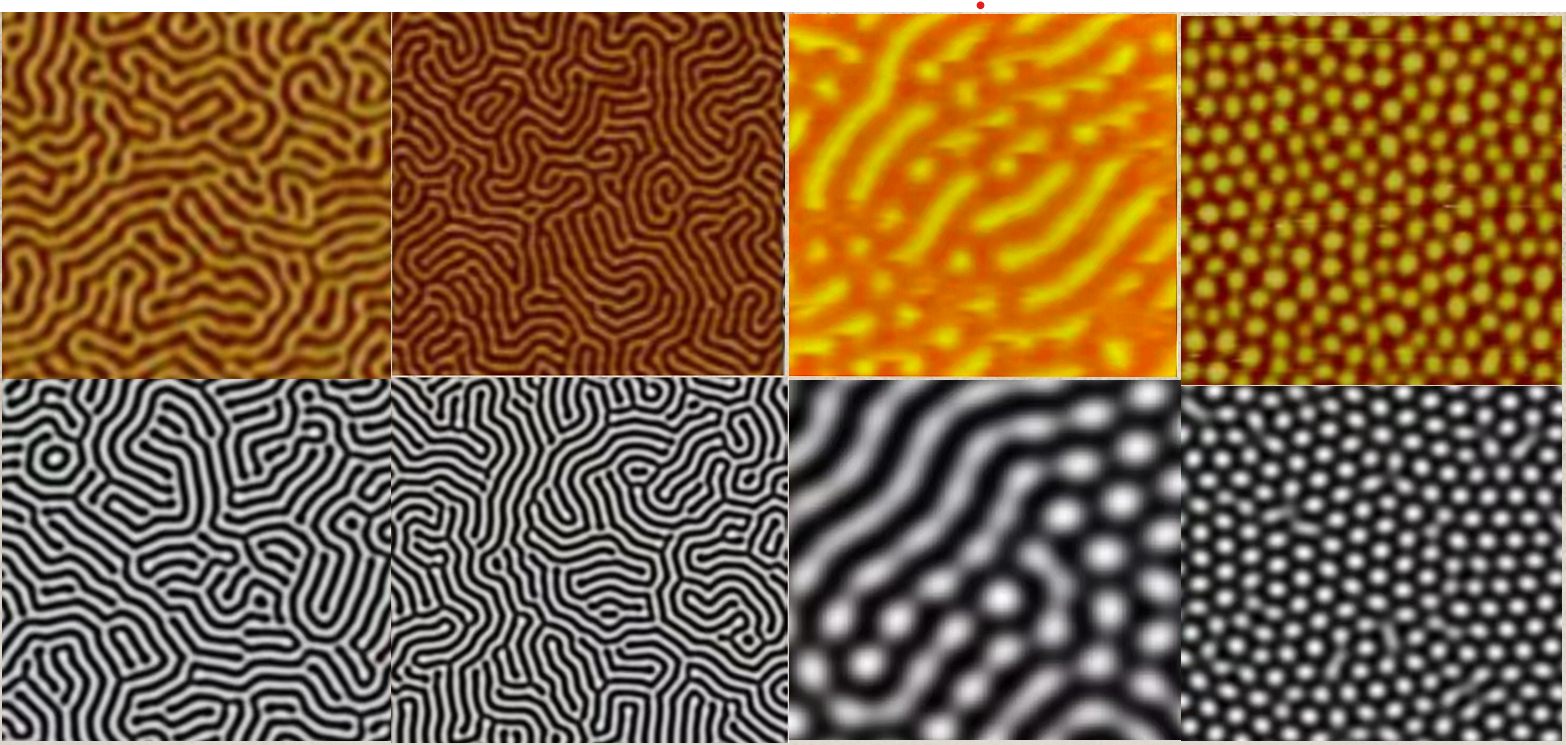
			}
			\caption{The images shown in red and yellow are MFM micrographs of PMA thin-film magnetic domains \cite{7,9,6}. The black-and-white images are maze-bubble phase diagrams simulated using the simulation model from Section 4.1 in MATLAB.}
		\end{figure}
		
		The steps in Section 4.1 can be expressed in code language, and using MATLAB for plotting, we can obtain an asymmetric phase diagram formed across the entire plane due to the varying differences between \( M_1 \) and \( M_2 \) at different positions. By modifying the parameters of equations (6) and (7), we can obtain the following images (Figure 4).

		A comparison reveals that, from a topological perspective, the images generated by the simulation model exhibit a high degree of similarity to the magnetic domain structures observed under the microscope. Thus, we can tentatively conclude that applying the LE model innovatively to electromagnetic simulations is not a far-fetched idea; interpreting the formation of magnetic domains through the lens of a reaction-diffusion system is indeed meaningful. (Note: The code used in this study can be viewed on GitHub at: 
		\href{https://github.com/PhoebeBai-l/Magnetic-Domain-Simulation-Code}{https://github.com/PhoebeBai-l/Magnetic-Domain-Simulation-Code.})
		
		\subsection{The case of magnetic domain under the influence of thickness and external field}
		
		According to \cite{6}, the topological structure change of $[Co/Pt]_N$ multilayer films in the remanent magnetization state depends on the thickness of the $Co$ layer and the applied magnetic field $H_m$.

		When the number of repetitions $N$ is 50, the thickness of the $Pt$ layer is 7 Å, the thickness of the $Co$ layer ranges from 0 to 60 Å, and the applied magnetic field $H_m$ ranges from 0 to $9 \times 10^4$ Oe, the topological structure of the $[Co/Pt]_N$ multilayer films primarily exhibits maze patterns, bubble patterns, and intermediate states between these two modes, as illustrated in Figure 5. These three topological patterns correspond to different $Co$ layer thicknesses and $H_m$ values, and the conditions for their formation can be clearly observed in the figure.
		\begin{figure}[ht]
			\centering
			\includegraphics[width=0.9\linewidth]{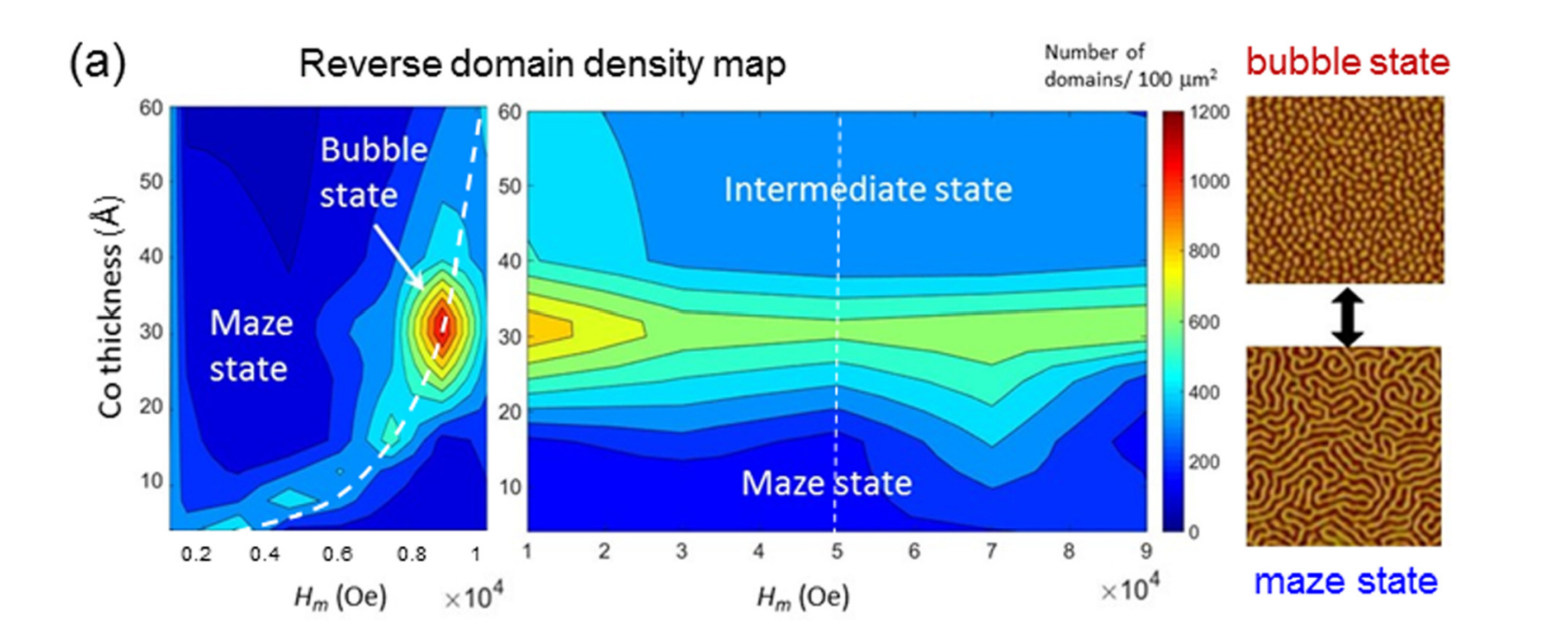}
			\caption{The topological structure change of PMA multilayers in the remanent magnetization state depends on the thickness of the film and the applied magnetic field\cite{6}.}
		\end{figure}\newpage
		\subsection{Simulation model refinement and analysis}
		
		To further refine the simulation model, we fix three parameters in the original system of equations as constants and assign physical meanings to the remaining two parameters. Parameter \(a\) represents the external field strength, and parameter \(w\) represents the material thickness. 
		\begin{align}
			\frac{\partial M_1}{\partial t} &= \Delta M_1 + a - M_1 - \frac{w M_1 M_2}{1 + M_1^2}, \\
			\frac{\partial M_2}{\partial t} &= 20 \left[ \Delta M_2 + 0.3 \left( M_1 - \frac{M_1 M_2}{1 + M_1^2} \right). \right]
		\end{align}
		In the context of the system of equations, parameter \(a\) corresponds to the initial growth rate of \(M_1\), which is similar to applying an external field strength to the PMA film. Parameter \(w\) controls the coupling reaction between the two variables, which is analogous to how the film thickness affects the exchange coupling of atomic magnetic moments in the vertical direction.

		The self-organizing phenomenon in reaction-diffusion systems arises from the coupling between chemical dynamics and diffusion processes. A stable spatial homogeneous state becomes unstable under certain conditions far from thermodynamic equilibrium. After undergoing a non-equilibrium phase transition, a series of spatiotemporal steady or dynamic patterns spontaneously emerge. From a mathematical perspective, the mechanism of this non-equilibrium phase transition can be divided into two categories. The first category originates from the instability of a constant spatial homogeneous solution in phase space. This is the main focus of our study and is also consistent with the physical understanding of the reasons behind magnetic domain formation. Such non-equilibrium phase transitions can be analyzed by perturbation theory of the steady-state solution to derive different types of spatiotemporal symmetry breaking \cite{123}. Through a combination of linear stability analysis and numerical verification, E. Mosekilde and F. Larsen plotted the phase diagrams of the original LE model ($w$=4) and its relationship with parameters \(a\) and \(b\) \cite{a}.
		
		\begin{figure}[ht]
			\centering
			\includegraphics[width=0.48\linewidth]{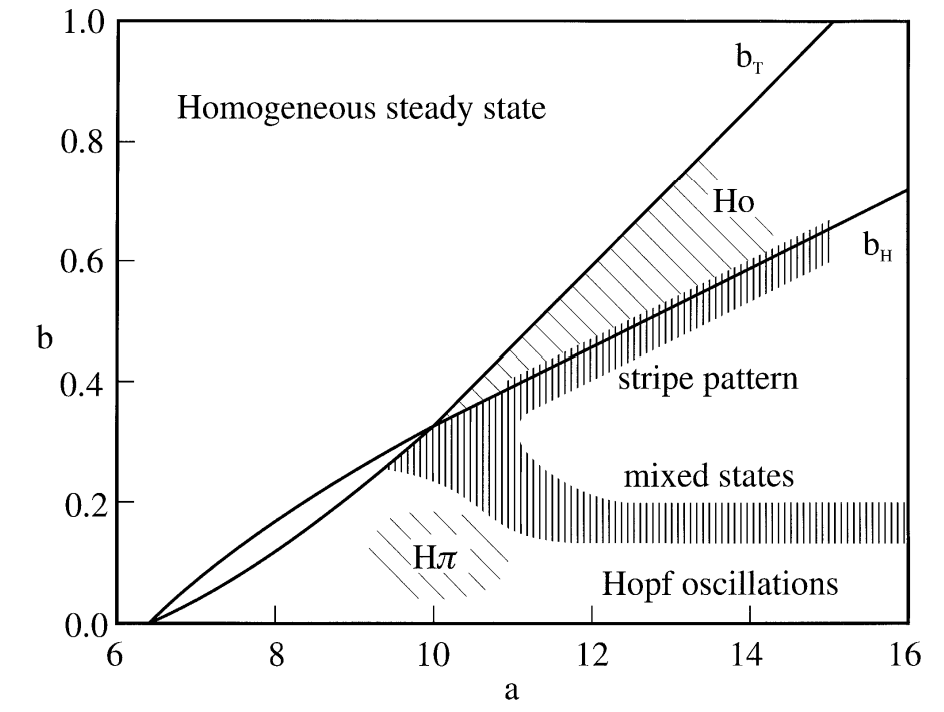}
			\caption{Phase diagram for the Lengyel-Epstein model as obtained from the linear stability analysis for \(c = 1.5\) and \(\delta = 11\)\cite{a}. Here, \( b_T \) represents the Turing bifurcation curve, and \( b_H \) represents the Hopf bifurcation curve.}
		\end{figure}

		In order to make the simulation results, namely the evolution of the maze and bubble phase diagrams, more consistent with the actual results, this section first attempts to determine the Hopf bifurcation and Turing bifurcation in the simulation model (with \( w \) as a variable) through linear stability analysis. Then, based on the results in the diagram, the approximate range of phase diagram shapes corresponding to different parameters is determined.
		
		Based on linear stability analysis\cite{123}, we first consider the equilibrium points of equations (8) and (9). For convenience of description, we write it as:
		\begin{align}
			\frac{\partial M_1}{\partial t}&= f(M_1, M_2,a,w)+\Delta M_1,
			\\
			\frac{\partial M_2}{\partial t} &=g(M_1, M_2)+20\Delta M_2.
		\end{align}
		Calculations show that the system has a unique equilibrium point \[(M_{10}, M_{20}) = (\alpha, \alpha^2 + 1),\] where \(\alpha = \frac{a}{w+1}\). 
		
		We apply a perturbation to this equilibrium point, letting \(M_1 = M_{10} + m_1\) and \(M_2 = M_{20} + m_2\), substitute into the equation, and expand the right-hand side of the equation in a Taylor series around the equilibrium point, discarding higher-order terms, to obtain: 
		\begin{align}
			\frac{\partial m_1}{\partial t} &=a_{11} m_1+ a_{12}  m_2 + \Delta m_1,
			\\
			\frac{\partial m_2}{\partial t} &= a_{21} m_1 + a_{22} m_2 +20 \Delta m_2.
		\end{align}
		Here,
		\begin{align}
			a_{11}&= \frac{(w-1)\alpha^2-w-1}{1+\alpha^2}, \\a_{12}&= -\frac{w\alpha}{1+\alpha^2},
			\\
			a_{21}&=\frac{40\alpha^{2} b }{1+\alpha^2}, \\a_{22}&=-\frac{20\alpha b}{1+\alpha^2}.
		\end{align}
		Based on the results in \cite{123}, the necessary condition for the Hopf bifurcation is:
		\begin{align}
			a_{11}+a_{22}=0,
		\end{align}
		and the necessary condition for the Turing bifurcation is:
		\begin{align}
			20a_{11}+a_{22}>2\sqrt{20(a_{11}a_{22}-a_{21}a_{12})}.
		\end{align}
		\newpage

		Therefore, we can plot the Hopf bifurcation curve and Turing bifurcation curve based on formulas (18) and (19).
		
		\begin{figure}[ht]
			\centering
			\includegraphics[width=0.74\linewidth]{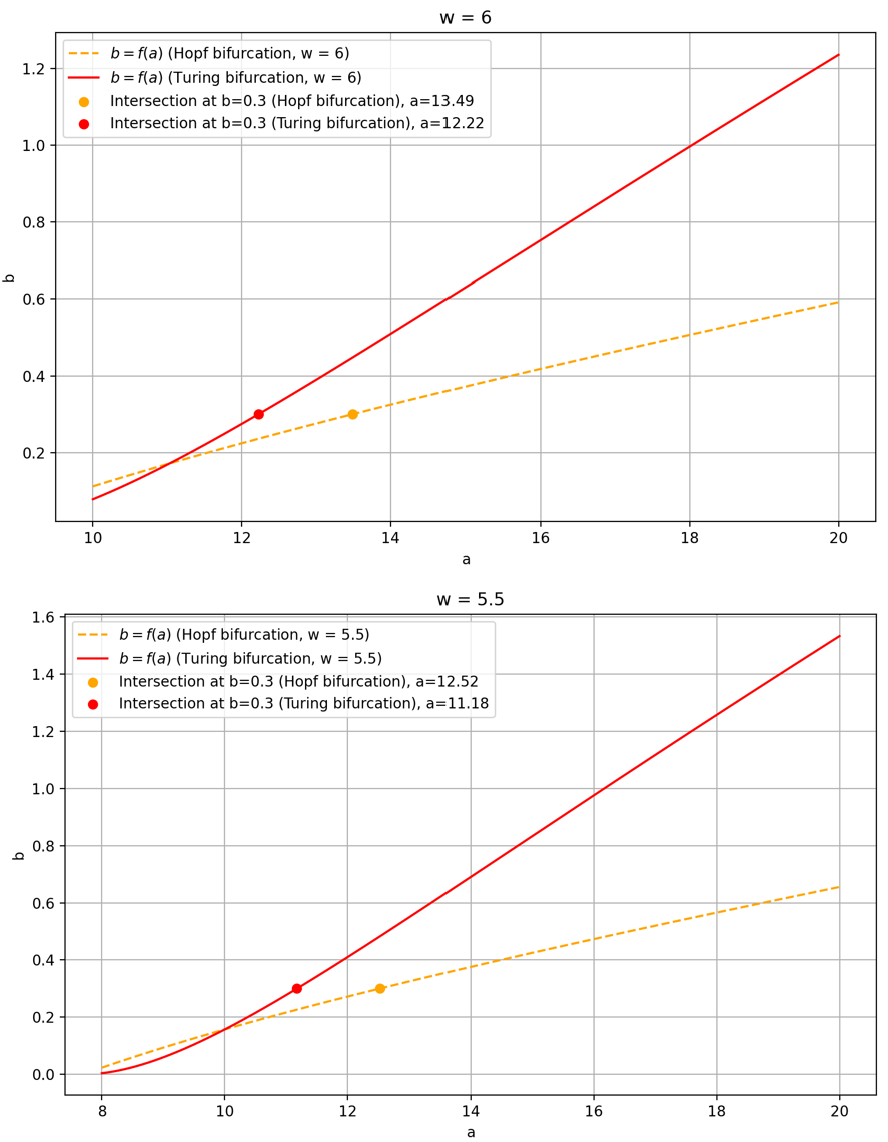}
			\caption{Turing bifurcation curve and Hopf bifurcation curve for the simulation model as obtained from the linear stability analysis for \(w = 6\) and \(w = 5.5\).}
		\end{figure}
		By comparing Figures 7 and 6, along with actual numerical simulations, we finally set the parameter ranges of the model to \( 4 < w < 6 \) and \( 12.4 < a < 13.9 \).
		\newpage
		\subsection{Simulation results in the case of Section 4.3 }
		
		After refining the simulation model, the next step is to adjust the parameters in the equations to simulate the effects of the Co layer thickness and the applied magnetic field on the magnetic domain structure. The main criterion for validating the simulation results is the result graph in Figure 5, which reflects the influence of $H_m$ and Co layer thickness on the magnetic domain pattern. It should be noted that the primary basis for judging whether the simulation results resemble actual images is the similarity in topological structure.

		Additionally, considering that the magnetic field applied in reality may not reach saturation, theoretically, the initial state of the square region should be adjusted. However, we believe that when the applied magnetic field is close to saturation, this effect can be represented by the value of $a$ in the system of equations. Therefore, no further adjustments are made to the initial states of $U$ and $V$.
		\begin{figure}[ht]
			\centering
			\includegraphics[width=0.6\linewidth]{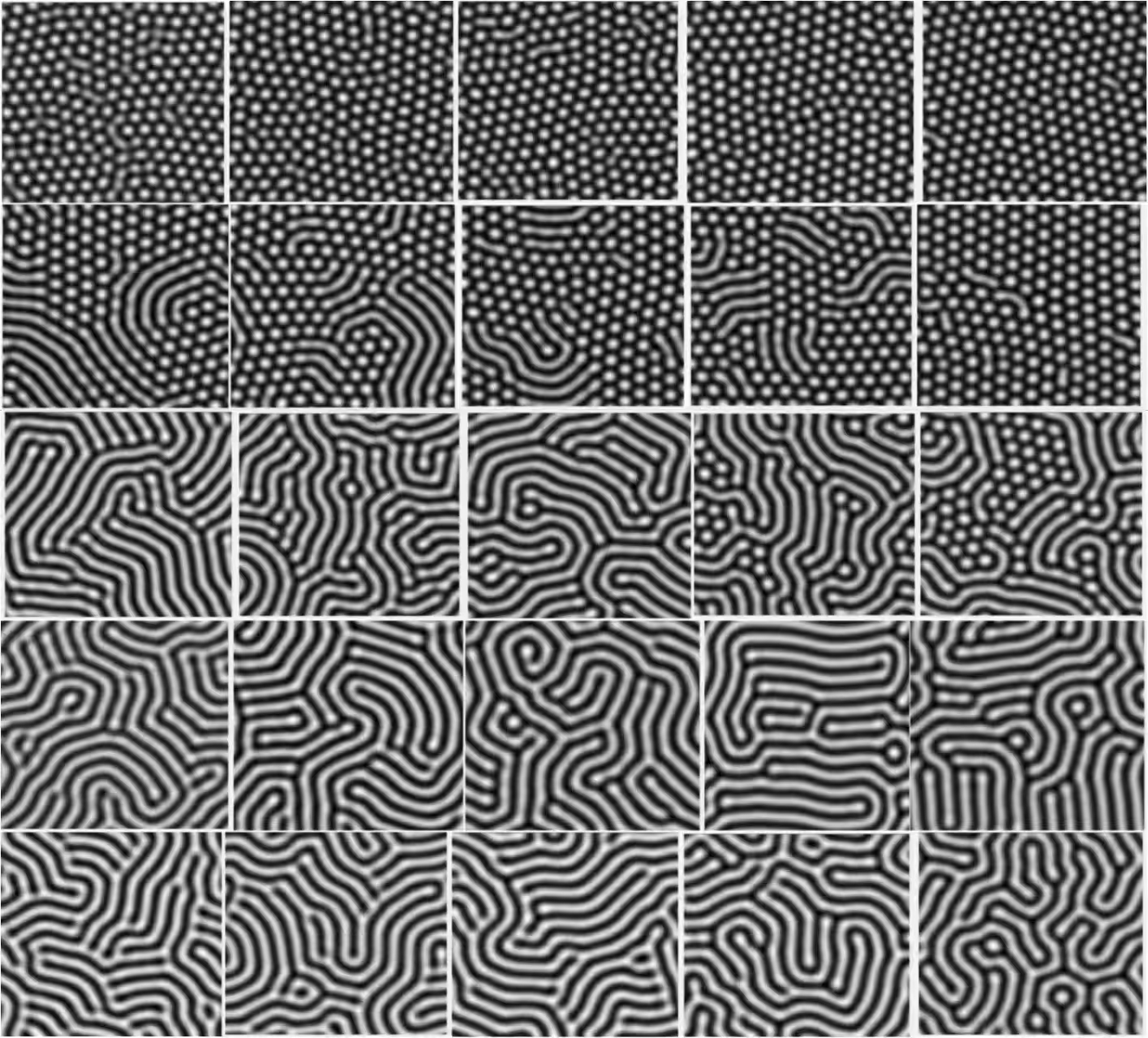}
			\caption{The simulation result}
		\end{figure}

		As shown in Figure 8, when the parameter $a$ varies between 12.4 and 13.9, and $w$ varies between 4 and 6, the phase diagram transitions from a labyrinth stripe phase diagram to a mixed labyrinth bubble phase diagram, and then to a bubble phase diagram. This is roughly consistent with the changes in the magnetic domain structure when the external field strength varies between 75\% and 95\% of the saturation magnetic field strength, and the material thickness varies between 10 and 30 Å.
		\subsection{Simulation conclusion}
		
		In summary, this simulation experiment successfully modeled the maze-bubble remanent magnetic domains of PMA films using the LE model, showing a high degree of similarity to actual results. The physical significance assigned to the parameters aligns well with real-world outcomes, and the model reproduces the maze-bubble domain patterns under different film thicknesses and external magnetic field influences. The experiment also demonstrates the feasibility of viewing magnetic domains from a reaction-diffusion perspective, showcasing the innovative application of the reaction-diffusion model in the field of electromagnetism.
		\section{Extensions}
		
		This section extends the content of this study by first introducing the cutting-edge magnetic storage device—the magnetic skyrmion, and demonstrating the strength of the simulation model through successful simulations of its morphology. This section also proposes another hypothesis based on the analogy between Turing patterns and magnetic domain patterns, drawing from the comparison between Turing patterns and magnetic domain patterns in the reaction-diffusion model.
		\subsection{Magnetic Skyrmion}
		
		Magnetic skyrmions are vortex-like magnetic structures with topologically protected properties, which appear in domain structures in thin films. Due to their nontrivial topological nature in real space, magnetic skyrmions exhibit rich and novel physical properties, such as the topological Hall effect. Due to their small size, high stability, and ease of manipulation, skyrmions have potential applications in high-density, low-energy storage devices\cite{z,x}.
		\begin{figure}[ht]
			\centering
			\includegraphics[width=0.5\linewidth]{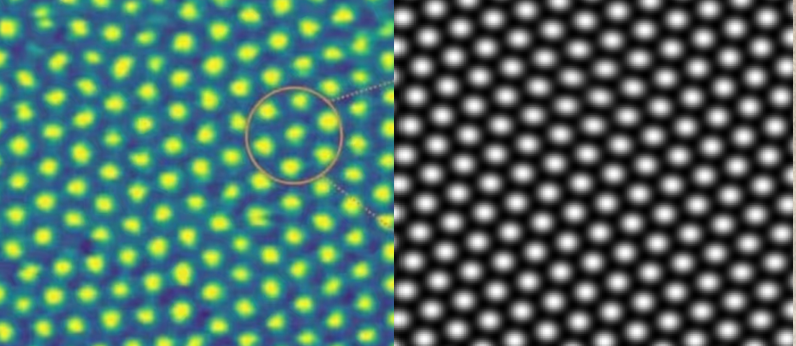}
			\caption{The left image\cite{y} shows a skyrmion under a magnetic force microscope, while the right image shows the result simulated using the model from section 4.1.}
		\end{figure}
		\begin{figure}[htbp]
			\centering
			\subfigure[The simulation result using the model from section 4.1]{
				\includegraphics[width=0.4\textwidth]{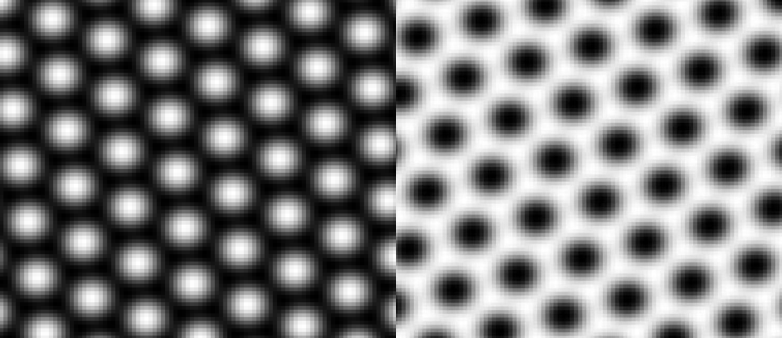}  
			}
			\hspace{0.05\textwidth}  
			\subfigure[Two different directional Skyrmions under Magnetic Force Microscopy (MFM)\cite{u}]{
				\includegraphics[width=0.4\textwidth]{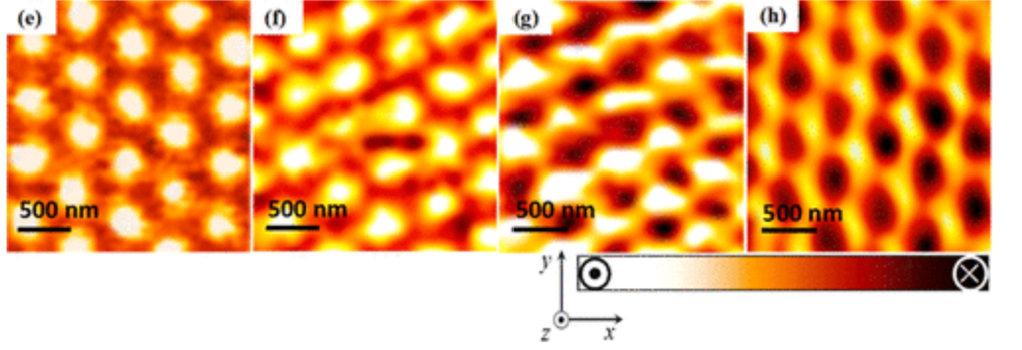}  
			}
			\caption{Comparison of simulation results with Skyrmions}
			\label{fig:side_by_side2}
		\end{figure}
		
		In fact, the bubble mode is essentially a magnetic skyrmion structure. By using the simulation model, we can simulate skyrmions. As illustrated in Figure 10, the simulation results are highly similar to the actual results. Moreover, by adjusting the parameters, two different skyrmion configurations with distinct magnetic moment directions can be obtained, which further demonstrates the power of the LE model and the feasibility of understanding complex microscopic magnetic behaviors from the perspective of reaction-diffusion systems.
		\subsection{Another hypothesis}
		
		In addition to the LE model, reaction-diffusion systems, such as the Gray-Scott model, can also produce Turing patterns. Figure 11 shows typical Turing patterns, whose topological structures bear a significant resemblance to the magnetic domain structures found in many magnetic materials. This similarity not only provides an alternative approach for simulating magnetic domain structures but also further validates the rationality of using the reaction-diffusion perspective to understand the formation of magnetic domains.
		
		\begin{figure}[ht]
			\centering
			\includegraphics[width=0.6\linewidth]{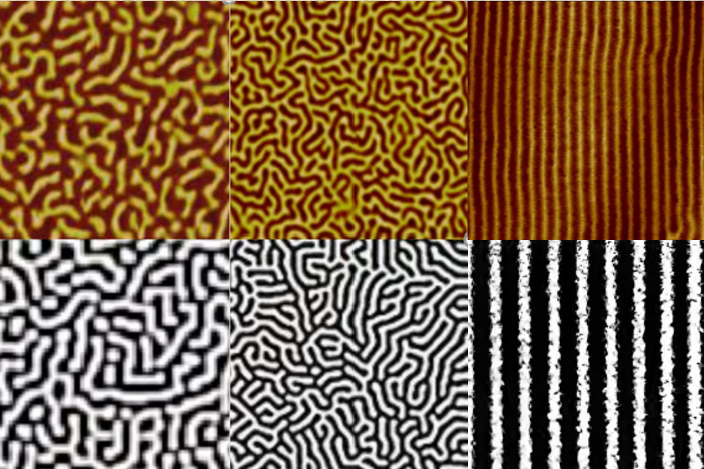}
			\caption{The images shown in red and yellow represent magnetic domain images of PMA films observed under an MFM microscope\cite{6,5.2.1,5.2.3}, while the black and white images are classic Turing patterns\cite{123,1234,12345}.}
		\end{figure}
		
		This similarity inspires us to form another hypothesis, magnetic domains are approximately equivalent to Turing patterns. The correctness of this hypothesis remains to be tested. Through the discussion in this article, we believe that the hypothesis holds in the case of PMA films. If this hypothesis is universally correct, it would have significant implications for the study of magnetic domains — we could draw parallels between the control theory and qualitative analysis in Turing patterns and magnetic domains. For example, we could apply the method used to control pattern morphology in Turing patterns, namely time-delayed feedback control (where a signal is extracted from a certain region of the system, delayed, and then fed back into the loop to couple with the system)\cite{17}, to the simulation of magnetic domains. This approach could be used to simulate the effects of temperature, external field strength, and scanning rate on magnetic domains.

		\section{Conclusions and Outlook}
		
		The advantages of this simulation experiment lie in the introduction of a clear system of equations to represent the formation of magnetic domains, and the innovative application of reaction-diffusion theory, which is commonly used in biochemistry, to electromagnetism. This approach holds significant value for the study of magnetic domains. However, the simulation results exhibit a relatively strong regularity in the images, which makes them appear less natural compared to actual results. Additionally, the study provides a qualitative analysis of magnetic domain formation but lacks a quantitative analysis, meaning that the simulation accuracy still needs improvement. 
		
		Considering improvements to this model, two possible approaches can be identified. One approach is to extend the simulation for uniaxial anisotropy. In this experiment, only films with vertical (uniaxial) anisotropy are simulated. For films with multi-axial anisotropy, multiple-component reaction-diffusion equations can be used for simulation. 
		\begin{align}
			\frac{\partial\mathbf{ q}}{\partial t} = D \Delta\mathbf{q}  + R(\mathbf{q}).
		\end{align}
		
		Here, $\mathbf{q}$ is a multivariable vector function, $D$ is the diffusion coefficient, and $R$ is the reaction function.
		
		Another method is to enhance the quantitative connection between the equations and the physical aspects. The parameters in the system of equations should have a closer and more rigorous connection to the actual physical quantities. At the same time, this experiment only simulates the effects of thickness and external field strength on the magnetic domain structure. For other influencing factors, parameters can be added to the system of equations or delayed feedback terms can be introduced.
		\begin{align}
			F=g_u[u(t-r)-u(t)].
		\end{align}
		
		Here, $F$ is a delayed feedback term, $g_u$ is a parameter, $r$ is a delayed time, $u$ is one of the variables.
		
		Looking ahead, by further researching and refining the reaction-diffusion model and incorporating the characteristics of real physical systems, we hope to more accurately simulate the evolution of magnetic domain structures. This direction not only helps deepen our understanding of the underlying mechanisms of magnetic materials but also promises to advance the application and development of self-organizing systems in other fields.
		
		\section{Declaration of generative AI in scientific writing}
		
		During the preparation of this work the author used GPT-4o in order to improve and polish language. After using this tool, the author reviewed and edited the content as needed and takes full responsibility for the content of the publication.

	\end{document}